\documentclass[12pt]{article}
\usepackage{graphicx}
\begin{document}
\renewcommand{\thefootnote}{\fnsymbol{footnote}}
\begin{titlepage}
\begin{flushright}
OU-HET 446 \\
hep-th/0306034
\end{flushright}

\vspace{10mm}
\begin{center}
{\Large\bf Chiral Bosons in Noncommutative Spacetime}
\vspace{25mm}

{\large
Yan-Gang Miao$^{a,}$\footnote{e-mail address: miao@het.phys.sci.osaka-u.ac.jp},
Harald J.W. M\"uller-Kirsten $^{b,}$\footnote{e-mail address: mueller1@physik.uni-kl.de}
and
Dae Kil Park$^{c,}$\footnote{e-mail address: dkpark@hep.kyungnam.ac.kr}}\\
\vspace{10mm}
${}^a$ {\em Department of Physics, Osaka University,
Toyonaka, Osaka 560-0043, Japan}

\vspace{4mm}
${}^b$ {\em Department of Physics, University of Kaiserslautern, P.O. Box 3049, D-67653 Kaiserslautern,
Germany}

\vspace{4mm}
${}^c$ {\em Department of Physics, Kyungnam University, Masan 631-701, Korea}
\end{center}

\vspace{12mm}
\centerline{{\bf{Abstract}}}
\vspace{5mm}

Underlying a general noncommutative algebra with both noncommutative coordinates and noncommutative momenta in a (1+1)-dimensional spacetime, a chiral boson Lagrangian with manifest Lorentz covariance is proposed by linearly imposing a generalized self-duality condition on a noncommutative generalization of massless real scalar fields. A significant property uncovered for noncommutative chiral bosons is that the left- and right-moving chiral scalars cannot be distinguished from each other, which originates from the noncommutativity of coordinates and momenta. An interesting result is that Dirac's method can be consistently applied to the constrained system whose Lagrangian explicitly contains space and time. The self-duality of the noncommutative chiral boson action does not exist.

\end{titlepage}
\newpage
\renewcommand{\thefootnote}{\arabic{footnote}}
\setcounter{footnote}{0}
\setcounter{page}{2}

\section{Introduction}

In recent years a large amount of literature on noncommutative field theory (NCFT)~\cite{s1} has appeared and much on noncommutative quantum mechanics (NCQM)~\cite{s2,s3} as well. The mathematical background of these theories is the noncommutative geometry~\cite{s4}. Simply speaking for the noncommutative Minkowski or Euclidean spacetime\footnote{There exist other compact manifolds, such as noncommutative spheres and tori, on which models of NCFT and NCQM are usually constructed.}, the algebra of coordinates of spacetime and their conjugate momenta satisfies the full Heisenberg commutation relation instead of the standard Heisenberg commutation relation. That is, the commutation relation of coordinates is no longer vanishing but either an antisymmetric constant or a tensor operator, while the commutation relation of momenta still keeps vanishing. The more general case is that the commutation relation of momenta is also non-vanishing, and such an algebra is called the general Heisenberg commutation relation here. Incidentally, comparing with a large amount of work based on the full Heisenberg commutation relation, there is only little work on NCFT~\cite{s5,s6,s7} and NCQM~\cite{s3} related to  momentum commutators of an antisymmetric tensor or simply of a constant magnetic field. This paper is devoted to the study of NCFT underlying the general algebra with both noncommutative coordinates and noncommutative momemta\footnote{The noncommutativity of momenta was considered mostly in order to incorporate an additional background magnetic field. It is, however, utilized in the present work to generalize chiral bosons from commutative to noncommutative (1+1)-dimensional flat spacetime.}.

As is well-known, the first published work on noncommutative spacetime was done by Snyder~\cite{s8} although the idea might be traced back earlier to W. Heisenberg, R.E. Peierls, W. Pauli, and J.R. Oppenheimer. Snyder's motivation to introduce the so-called quantized spacetime was to remove the divergence trouble caused by point interactions between matter and fields. As is now accepted widely, instead of assuming a discrete spacetime theorists have developed a sufficiently
good renormalized theory to overcome this difficulty in the spacetime with scale larger than Planck's. However, the recent motivation to study the NCFT and NCQM originates, as has been recognized extensively, from the intimate relationship between NCFT and M-theory and string theory~\cite{s9} and between NCFT and the quantum Hall effect~\cite{s10}, respectively. Namely, some low-energy effective theory of open strings with a nontrivial background can be described by NCFT and thus some relative features of string theory may be clarified through the NCFT within a framework of quantum field theory, and on the other hand the quantum Hall effect can be deduced from the Abelian noncommutative Chern-Simons theory at level $n$ shown to be exactly equivalent to the Laughlin theory at filling fraction $1/n$.

The noncommutativity of spacetime can be described by the way of Weyl operators or that of normal functions with a suitable definition of star-products. The relationship between these is, as was stated~\cite{s1}, that the noncommutativity of spacetime may be encoded through ordinary products in the noncommutative $C^{\ast}$-algebra of Weyl operators, or equivalently through the deformation of the product of the commutative $C^{\ast}$-algebra of functions on spacetime to a noncommutative star-product. If coordinates and their conjugate momenta satisfy the full Heisenberg commutation relation and the coordinate commutator is an antisymmetric constant tensor, the star-product is just the Moyal-product~\cite{s11}; if, in general, the commutators among coordinates and momenta are operator functions of coordinates and momenta themselves, the relationship of the two ways of description mentioned above may maintain, but the corresponding star-product requires a more complicated formula~\cite{s5}. One important result deduced from the definition of the Moyal-product is that the integration of the Moyal-product of two normal functions equals that of the ordinary product of the two functions in any dimensional Minkowski or Euclidean spacetime. In physics quadratic terms of actions remain unchanged when coordinates of spacetime are shifted from commutative to noncommutative\footnote{In this paper we only generalize the full Heisenberg commutation relation to which a constant commutator of momenta is added. This is permitted because the commutation relation of momenta is independent of that of coordinates.}. As a consequence, no noncommutative generalization exists for chiral bosons with quadratic Lagrangians if the noncommutative algebra is restricted to the full Heisenberg commutation relation. We deviate from the discussion of noncommutativity temporally and give a brief introduction of chiral {\em p}-forms in an ordinary (commutative) $2(p+1)$-dimensional Minkowski spacetime\footnote{For the $p=0$ case, a chiral $0$-form in (1+1)-dimensional spacetime is usually called a chiral boson which describes a left- or right-moving boson in one spatial dimension. We focus on chiral bosons in this paper. The generalization to chiral {\em p}-forms ($p\geq 1$) will be considered separately.}.

Chiral {\em p}-forms, described by an antisymmetric {\em p}th order tensor, exist in $2(p+1)$-dimensional spacetime since their
external derivatives, {\em i.e.}, the field strengths, are ({\em p}+1)-forms
and the Hodge duals as well, which requires the spacetime dimensions to be
twice that of the form number of the field strengths. The chirality usually
means that the field strengths satisfy a self-duality condition, that is, roughly speaking\footnote{In fact the self-duality condition takes a different form when {\em p} is odd or even. For details, see Ref.~\cite{s15}.}, the field strengths equal to their Hodge duals. 
The main reason why chiral {\em p}-forms receive much attention
is that they appear in various theoretical models that relate to
superstring theories, and reflect especially the existence of a variety of
important dualities that connect these theories among one another.
One has to envisage two basic problems in a Lagrangian description of 
chiral {\em p}-forms: one is the consistent quantization and the other is the
harmonic combination of manifest duality and Lorentz covariance, since the
equation of motion of a chiral {\em p}-form, {\em i.e.}, the
self-duality condition, is first order with respect to the derivatives of
space and time. In order to solve these problems, some non-manifestly Lorentz
covariant~\cite{s12} and manifestly Lorentz covariant~\cite{s13,s14,s15} models have
been proposed. 
It is remarkable that these chiral {\em p}-form models have close 
relationships 
among one another, especially various dualities that have been
demonstrated in detail from the points of view of both configuration~\cite{s15,s16}
and momentum~\cite{s17} spaces. Among the models mentioned above we are interested in the so-called linear constraint formulation of chiral bosons~\cite{s14} where the self-duality itself, instead of the square proposed by Siegel~\cite{s13}, is imposed on a massless real scalar field. Although it has some defects~\cite{s18}, the linear formulation strictly describes a chiral boson from the point of view of equations of motion at both the classical and quantum levels\footnote{Another consideration to choose the linear model is that its Lagrangian is quadratic for which there is no noncommutative generalization based on the full Heisenberg commutation relation.}.

In this paper we propose a noncommutative formulation of the linear model of chiral bosons underlying the noncommutative algebra with a constant commutator of coordinates and simultaneously with a constant commutator of momenta as well. Completely following the procedure in ordinary spacetime, we first suggest a noncommutative generalization of massless real scalar fields and then separate it into its corresponding ``left-'' and ``right-moving'' scalars. The equation of motion for the ``left-'' or ``right-moving'' boson is, in fact, the generalized self-duality condition or chirality. Secondly, introducing a Lagrange multiplier we impose the self-duality condition upon the noncommutative real scalar field in a manifestly Lorentz covariant way. Solving the equation of motion under the requirement of reality of the scalar field we find that the solutions indeed describe one left- or right-moving object but they can convert from left- to right-moving, and {\em vice versa}. That is, we obtain some ``fuzzy'' spatial dimension, which should show the noncommutativity of coordinates and momenta.
Furthermore, the noncommutative model of chiral bosons is quantized by using the Dirac method~\cite{s19}, and it is found that the method is applicable to the model that explicitly contains space and time. In addition, we are also interested in the duality symmetry of the noncommutative chiral boson action and make a brief discussion in a similar way to that of the commutative case.

The arrangement of this paper is as follows. In the next section, we first write out the noncommutative algebra as our starting point, and then search for a suitable realization\footnote{Here the realization means a linear mapping from the noncommutative algebra to the commutative algebra where corresponding coordinates and conjugate momenta satisfy the standard Heisenberg commutation relation.}. In general, realizations of a noncommutative algebra are not unique and not all the realizations can be utilized to deduce a physically accepted noncommutative generalization. We find such a realization that meets our requirement and then construct the equation of motion for noncommutative massless real scalar fields. By using light-cone coordinates we can therefore obtain the corresponding Lagrangian that takes the form of a normal product of the generalized ``left-'' and ``right-handed'' chiralities. Moreover, this system is also analyzed briefly by using the canonical Hamiltonian quantization. In section 3, following the procedure suggested in Ref.~\cite{s14} we propose the Lagrangian with the manifest Lorentz covariance for noncommutative chiral bosons. We emphasize that the requirement of reality of the scalar field plays a crucial role {\em not only} in solving the equation of motion {\em but also} in establishing a consistent theory. From this requirement we derive a constraint or condition that two light-cone coordinates should satisfy, which looks quite like a quantum condition in quantum mechanics, and from this constraint we deduce the ``fuzzy'' phenomenon mentioned above. In section 4, by utilizing the Dirac method we quantize the noncommutative linear model of chiral bosons and find that the quantized theory is a quite natural generalization of that of commutative chiral bosons. In particular, we prove that the model possesses ${\cal PT}$ symmetry~\cite{s20} and therefore survives although the Hermiticity is not maintained. Because of the interest in duality symmetries for ordinary chiral {\em p}-form (including chiral boson) actions we then in section 5 turn to the discussion in this aspect for the noncommutative model constructed in section 3, and discover, however, that the model is not self-dual for a finite noncommutative parameter, which is different from the case in the commutative spacetime. Finally, we conclude our findings in section 6.

The notation we use throughout this paper is as follows:
\begin{equation}
{\eta}^{{\mu}{\nu}}={\rm diag}(-1,+1),
\end{equation}
stands for the flat metric of the (1+1)-dimensional Minkowski spacetime.
Greek indices (${\mu},{\nu},{\sigma},\cdots$) run over 0,1. The completely antisymmetric tensor takes the form
\begin{equation}
{\epsilon}^{01}=-{\epsilon}^{10}=+1.
\end{equation}
Moreover, for the sake of convenience of using light-cone coordinates in (1+1)-dimensional spacetime, we define
\begin{equation}
x^{+} \equiv \frac{1}{\sqrt{2}}(x^{0}+x^{1}),\hspace{10mm}
x^{-} \equiv \frac{1}{\sqrt{2}}(x^{0}-x^{1}),
\end{equation}
and obtain their derivatives as
\begin{equation}
{\partial}_{+}=\frac{1}{\sqrt{2}}({\partial}_{0}+{\partial}_{1}),\hspace{10mm}
{\partial}_{-}=\frac{1}{\sqrt{2}}({\partial}_{0}-{\partial}_{1}).
\end{equation}
Correspondingly, the metric takes the form
\begin{equation}
{\eta}^{++}={\eta}^{--}=0,\hspace{10mm}
{\eta}^{+-}={\eta}^{-+}=-1. 
\end{equation}

\section{Noncommutative generalization for massless real scalar fields}

The noncommutative algebra that is dealt with as our starting point takes the form\footnote{In this section some discussions are also applicable to $D$-dimensional spacetime if no limitation is added.}
\begin{equation}
[{\hat{x}}^{\mu},{\hat{x}}^{\nu}]=i{\theta}^{{\mu}{\nu}},\hspace{10mm}
[{\hat{x}}^{\mu},{\hat{p}}_{\nu}]=i{\delta}^{\mu}_{\nu},\hspace{10mm}
[{\hat{p}}_{\mu},{\hat{p}}_{\nu}]=i{\omega}_{{\mu}{\nu}},
\end{equation}
where the Planck constant $\hbar$ is set to one, and ${\theta}^{{\mu}{\nu}}$ and ${\omega}_{{\mu}{\nu}}$ are two independent real constant tensors\footnote{As explained in section 1, requiring a constant ${\theta}^{{\mu}{\nu}}$ the star-product reduces to the Moyal-product. For simplicity, we introduce a non-zero constant commutator of momenta in order to establish a noncommutative formulation for quadratic Lagrangian theories.}. If ${\omega}_{{\mu}{\nu}}=0$ but ${\theta}^{{\mu}{\nu}}\neq 0$, the algebra (6) reduces to the full Heisenberg commutation relation. In Ref.~\cite{s6} a special choice, ${\omega}_{{\mu}{\nu}}=-({\theta}^{-1})_{{\mu}{\nu}}$, was considered. Next we look for such a realization of the noncommutative algebra (6), $(x^{\mu},p_{\mu})$, with the property that we can construct a noncommutative generalization of massless real scalar fields. Although realizations are not unique, they all satisfy the standard Heisenberg commutation relation (see footnote 7):
\begin{equation}
[{x}^{\mu},{x}^{\nu}]=0,\hspace{10mm}
[{x}^{\mu},{p}_{\nu}]=i{\delta}^{\mu}_{\nu},\hspace{10mm}
[{p}_{\mu},{p}_{\nu}]=0,
\end{equation}
where ${p}_{\mu}=-i{\partial}/{\partial}x^{\mu}$. We may write out a realization with independent commutators of coordinates and momenta, for example, ${{\hat x}^{\mu}}=x^{\mu}-\frac{\theta}{2}({\eta}^{{\mu}{\nu}}+{\epsilon}^{{\mu}{\nu}})p_{\nu}$, ${{\hat p}_{\mu}}=p_{\mu}-\frac{\omega}{2}({\eta}_{{\mu}{\nu}}+{\epsilon}_{{\mu}{\nu}})x^{\nu}$, where ${\theta}^{01}=\theta$ and ${\omega}_{01}=\omega$. However, this is not the one we desire. After making many trials, we find that one suitable realization is just the (1+1)-dimensional case of Ref.~\cite{s6}\footnote{For a general $D$-dimensional case, the realization of eq.(6) with ${\omega}_{{\mu}{\nu}}=-({\theta}^{-1})_{{\mu}{\nu}}$ takes the form: ${\hat{x}}^{\mu}=\frac{1}{2}{x}^{\mu}-{\theta}^{{\mu}{\nu}}p_{\nu}$ and ${\hat{p}}_{\mu}=p_{\mu}-\frac{1}{2}({\theta}^{-1})_{{\mu}{\nu}}x^{\nu}$, where Greek indices run over $0,1,\cdots,D-1$.} that possesses the property mentioned above
\begin{equation}
{\hat{x}}^{\mu}=\frac{1}{2}{x}^{\mu}-{\theta}{\epsilon}^{{\mu}{\nu}}p_{\nu},\hspace{10mm}
{\hat{p}}_{\mu}=p_{\mu}-\frac{1}{2}{\theta}^{-1}{\epsilon}_{{\mu}{\nu}}x^{\nu}.
\end{equation}
The feature of this realization is that the noncommutativity of coordinates and that of momenta are not independent, {\em i.e.}, they are {\em mixed}, which is obvious because of the relation ${\hat{x}}^{\mu}=-{\theta}^{{\mu}{\nu}}{\hat{p}}_{\nu}$, where ${\theta}^{{\mu}{\nu}}={\theta}{\epsilon}^{{\mu}{\nu}}$ for our case. This mixture will, of course, appear in equations of motion and Lagrangians proposed below. For the sake of explicitness in later discussion, we rewrite the noncommutative algebra\footnote{The commutation relation in noncommutative 1+1 dimensions (9) does not distinguish between space and time. Thus ${\hat x}^0$ is treated or ``quantized'' like a space coordinate. Space and time are on the equal level,
which is the feature of the (1+1)-dimensional spacetime. In higher (than two) dimensions, the noncommutativity of space is described by ${\theta}_{ij}$, where {$i,j,\cdots$} run over $1,2,\cdots,D-1$, but the noncommutativity of time is described by ${\theta}_{0i}$. Therefore, in higher dimensions
one may consider the spacetime with only noncommutative space by setting
${\theta}_{0i}$ to zero. However, in 1+1 case, there is only one independent
component of ${\theta}^{{\mu}{\nu}}$, and thus one cannot make a similar treatment.} we focus on in this paper
\begin{equation}
[{\hat{x}}^{0},{\hat{x}}^{1}]=i{\theta},\hspace{10mm}
[{\hat{x}}^{\mu},{\hat{p}}_{\nu}]=i{\delta}^{\mu}_{\nu},\hspace{10mm}
[{\hat{p}}_{0},{\hat{p}}_{1}]=i{\theta}^{-1}.
\end{equation}

Let us recall briefly the procedure of establishing field theory from mechanics in the ordinary spacetime. We start with the energy-momentum relation, $P^{\mu}P_{\mu}+m^2=0$, where $P_{\mu}$ stands for the energy and momentum of a relativistic particle and {\em m} its rest mass. Replacing $P_{\mu}$ by $p_{\mu}$, where the canonical pair $(x^{\mu},p_{\mu})$ satisfies the standard Heisenberg commutation relation (7), and multiplying a scalar field ${\Phi}(x)$ on both sides of the energy-momentum relation, we then obtain the equation of motion for complex scalar fields, $(p^{\mu}p_{\mu}+m^2){\Phi}(x)=0$, which describes a massive and charged boson. This procedure was directly utilized~\cite{s6} through replacing $p_{\mu}$ by ${\hat p}_{\mu}$ to derive the equation of motion for noncommutative complex scalar fields,
\begin{equation}
({\hat p}^{\mu}{\hat p}_{\mu}+m^2){\Phi}(x)=0,
\end{equation}
where ${\hat p}_{\mu}$ takes the realization mentioned in footnote 10. In order to arrive at our aim, we first need a noncommutative generalization for massless real scalar fields. To this end, we set $m=0$ in eq.(10) and obtain possible generalizations in three different ways: (i) ${\hat p}^{\mu\ast}{\hat p}_{\mu}\phi(x)=0$, (ii) $\frac{1}{2}({\hat p}^{\mu}{\hat p}_{\mu}+{\hat p}^{\mu\ast}{\hat p}^{\ast}_{\mu})\phi(x)=0$, and (iii) $\frac{i}{2}({\hat p}^{\mu}{\hat p}_{\mu}-{\hat p}^{\mu\ast}{\hat p}^{\ast}_{\mu})\phi(x)=0$, where $\ast$ means complex conjugate and $\phi(x)$ is a real scalar field, ${\phi}^{\ast}(x)=\phi(x)$. In terms of the realization (8) in (1+1)-dimensional spacetime case (i) corresponds to the equation of motion
\begin{equation}
{\partial}^{\mu}{\partial}_{\mu}{\phi}-\frac{1}{4{\theta}^2}x^{\mu}x_{\mu}\phi=0.
\end{equation}
This is the noncommutative generalization that we expect.
Obviously, it has the Lorentz covariance. When ${\theta}$ goes to infinity, {\em i.e.}, ${\hat x}^{0}$ and ${\hat x}^{1}$ tend to infinite large noncommutative but ${\hat p}_{0}$ and ${\hat p}_{1}$ to commutative, eq.(11) reduces to the ordinary equation of motion of massless real scalar fields, ${\partial}^{\mu}{\partial}_{\mu}{\phi}=0$. See further discussion of Lagrangians in the following. Note that case (ii) and case (iii), based on the realization (8), correspond to the equations of motion,         
${\partial}^{\mu}{\partial}_{\mu}{\phi}+\frac{1}{4{\theta}^2}x^{\mu}x_{\mu}\phi=0$ and
${\epsilon}^{{\mu}{\nu}}x_{\mu}{\partial}_{\nu}\phi=0$, respectively. Case (ii) seems to be another suitable generalization, its corresponding Lagrangian, however, cannot be separated to the product of generalized ``left-'' and ``right-handed'' chiralities (see next paragraph for details) and therefore cannot be used to construct a Lorentz covariant linear model of noncommutative chiral bosons. As to case (iii) the equation of motion is first order and is not a natural generalization of the ordinary equation of motion. As a consequence, we only choose case (i) to fulfil our purpose.

In light-cone coordinates eq.(11) can be rewritten as
\begin{equation}
{\partial}_{+}{\partial}_{-}{\phi}-\frac{1}{4{\theta}^2}x_{+}x_{-}\phi=0.
\end{equation}
We may construct the following Lagrangian
\begin{equation}
{\cal L}=-D_{+}{\phi}D_{-}{\phi}, 
\end{equation}
where $D_{+}$ and $D_{-}$ are defined by
\begin{eqnarray}
D_{+}& \equiv & -i{\partial}_{+}+\frac{1}{2\theta}x_{+},\nonumber \\
D_{-}& \equiv & -i{\partial}_{-}-\frac{1}{2\theta}x_{-}.
\end{eqnarray}
By means of the variation principle we derive the equation of motion as follows:
\begin{equation}
({D_{+}}^{\ast}D_{-}+D_{+}{D_{-}}^{\ast}){\phi}=0,
\end{equation}
which is, of course, exactly the same as eqs.(11) and (12) but in a different formulation that will be useful in the next section. It is evident that when ${\theta}$ goes to infinity the Lagrangian (13) reduces to the form, ${\cal L}_{0}=-{\partial}_{+}{\phi}{\partial}_{-}{\phi}$, that is, when the noncommutative algebra (9) reduces to the full Heisenberg commutation relation the Lagrangian of noncommutative generalization of massless real scalar fields simplifies to that of the ordinary spacetime. The reason is, as explained in section 1, that the Moyal-product, though with very large noncommutativity of coordinates in our case, satisfies in particular 
\begin{equation}
\int d^{2}x D_{+}{\phi}(x){\star}D_{-}{\phi}(x)=\int d^{2}x D_{+}{\phi}(x)D_{-}{\phi}(x),
\end{equation}
where $\star$ stands for the Moyal-product.
As expected, eq.(13) takes the form of the product of $D_{-}{\phi}$ and $D_{+}{\phi}$ which are called generalized ``left-'' and ``right-handed'' chiralities\footnote{For the equation of motion of case (ii), ${\partial}_{+}{\partial}_{-}{\phi}+\frac{{\omega}^2}{4}x_{+}x_{-}\phi=0$, the corresponding Lagrangian may be written as: ${\cal L}^{\prime}=-(-i{\partial}_{+}-\frac{\omega}{2}x_{-}){\phi}(-i{\partial}_{-}-\frac{\omega}{2}x_{+}){\phi}$, which is Lorentz covariant if total derivatives are dropped. Evidently, it is not the product of two chiralities. This is the reason why the case was not selected.}, respectively.

Now let us turn to a brief discussion of the Hamiltonian quantization for the system described by eq.(13). For the sake of convenience, we rewrite eq.(13) as follows:  
\begin{equation}
{\cal L}=\frac{1}{2}({\partial}_{0}{\phi})^{2}-\frac{1}{2}({\partial}_{1}{\phi})^{2}-\frac{1}{8{\theta}^2}[-(x^0)^2+(x^1)^2]{\phi}^2,
\end{equation}
where total derivatives have been dropped. Note that it contains explicitly space and time. The canonical momentum conjugate to $\phi$ is defined by
\begin{equation}
{\pi}_{\phi} \equiv {\partial}{\cal L}/{\partial}({\partial}_{0}\phi)={\partial}_{0}{\phi},
\end{equation}
which gives no primary constraints.
In terms of the Legendre transformation we obtain the canonical Hamiltonian
\begin{equation}
{\cal H}=\frac{1}{2}[{{\pi}_{\phi}}^2+({\partial}_{1}{\phi})^2]+\frac{1}{8{\theta}^2}[-({x^0})^2+({x^1})^2]{\phi}^2.
\end{equation}
In commutative spacetime the Hamiltonian contains only the first term and is, of course, positive definite. However, in noncommutative spacetime the Hamiltonian is no longer positive definite because of the appearance of a finite noncommutative parameter. Fortunately, this defect does not relate to our following study of constructing a noncommutative generalization of the linear constraint model of chiral bosons. In fact, the model is not positive definite even in ordinary spacetime~\cite{s14}. Nevertheless, the merit of the model, if compared with other chiral boson theories, includes that its quantized theory does not show any violation of causality, besides the merit of the manifest Lorentz covariance and precise description of one direction moving object at both the classical and quantum levels. 
By using the equal-time commutation relation
\begin{equation}
{[{\phi}(x),{\pi}_{\phi}(y)]}_{ET}=i{\delta}(x^1-y^1),\hspace{6mm}
{[{\phi}(x),{\phi}(y)]}_{ET}=0={[{\pi}_{\phi}(x),{\pi}_{\phi}(y)]}_{ET},
\end{equation}
we then arrive at the first-order Hamiltonian equations of motion
\begin{eqnarray}
{\partial}_{0}{\phi}(x)&=&-i\int dy^{1}{[{\phi}(x),{\cal H}(y)]}_{ET}={\pi}_{\phi},\nonumber \\
{\partial}_{0}{\pi}_{\phi}(x)&=&-i\int dy^{1}{[{\pi}_{\phi}(x),{\cal H}(y)]}_{ET}={{\partial}_{1}}^2{\phi}-\frac{1}{4{\theta}^2}[-(x^0)^2+(x^1)^2]{\phi}.
\end{eqnarray}
We can easily derive eq.(11) or eq.(12) from the Hamiltonian equations of motion if eliminating ${\pi}_{\phi}$. This shows that the Hamiltonian quantization is applicable to the system whose Lagrangian (17) explicitly contains space and time.

\section{Noncommutative linear constraint model of chiral bosons}

Introducing an auxiliary vector field, ${\lambda}_{\mu}(x)$, and linearly imposing in a Lorentz covariant way the left-handed chirality, $D_{-}{\phi}$, upon the noncommutative generalization of massless real scalar fields described by eq.(13), we propose a noncommutative linear constraint model of chiral bosons defined by
\begin{equation}
{\cal L}_{c}=-D_{+}{\phi}D_{-}{\phi}-{\lambda}_{+}D_{-}{\phi}, 
\end{equation}
where ${\lambda}_{+} \equiv \frac{1}{\sqrt{2}}({\lambda}_{0}+{\lambda}_{1})$. Variations with respect to ${\lambda}_{+}$ and $\phi$, respectively, lead to the equations of motion
\begin{equation}
D_{-}{\phi}=0,
\end{equation}
and
\begin{equation}
({D_{+}}^{\ast}D_{-}+D_{+}{D_{-}}^{\ast}){\phi}+{D_{-}}^{\ast}{\lambda}_{+}=0.
\end{equation}
Obviously, eq.(23) plays a similar role to ${\partial}_{-}{\phi}=0$ in ordinary spacetime. This can be verified immediately by solving the equation.   

In general, the solution of eq.(23) takes the form
\begin{equation}
{\phi}_{NC}^{(-)}=f(x^{+}){\rm exp}(-i\frac{1}{2\theta}x^{+}x^{-}), 
\end{equation}
where $f(x^{+})$ is an arbitrary real function of $x^{+}$. We emphasize that it is the phase factor that leads to the fuzzy phenomenon of left- and right-movings. At first we remind that $\phi$ is {\em real} and thus the solution (25) should be real, too. Secondly, in order to establish a consistent theory we have to impose the requirement of reality on the solution. As a consequence, we obtain the following constraint or condition that $x^{+}$ and $x^{-}$ have to satisfy
\begin{equation}
x^{+}x^{-}=2{\theta}n{\pi},  
\end{equation}
where $n$ is an integer\footnote{$n=0$ is a special case which describes chiral bosons moving on the light-cone defined by $x^{+}x^{-}=0$. For this case $x^{+}$ and $x^{-}$ take values on their corresponding axes, respectively. The left- and right-moving degrees of freedom cannot convert into each other because only one of them exists moving in one direction on the light-cone. As a result, this case corresponds to that in the commutative spacetime.} and called ``quantum number''. It looks quite like a quantum condition as we pointed out before. With this condition eq.(25) reduces to 
\begin{equation}
{\phi}_{NC}^{(-)}=f(x^{+}), 
\end{equation}
where a minus sign has been absorbed when $n$ is odd. On the one hand, eq.(27) indeed describes a left-moving chiral boson, but on the other hand it also describes a right-moving chiral boson, which can be seen clearly if we substitute $x^{+}=2{\theta}n{\pi}/x^{-}$, {\em i.e.}, the quantum condition into eq.(27) and derive the following formulation of the solution
\begin{equation}
{\phi}_{NC}^{(-)}=f(2{\theta}n{\pi}/x^{-}) \equiv g(x^{-}), 
\end{equation}
where $g(x^{-})$ is an arbitrary real function of $x^{-}$. Eq.(28) coincides with the solution of the equation of motion satisfied by a right-moving chiral boson, that is, $D_{+}{\phi}=0$ together with the requirement of reality. Consequently, the left- and right-moving chiral bosons convert into each other. Namely, they cannot be distinguished from each other. This phenomenon originates from the noncommutativity of coordinates and momenta (see eq.(9)) and appears typically in noncommutative chiral bosons. We may call it the fuzzy spatial dimension (ambiguity of the direction of the spatial dimension $x^1$) or fuzzy left- and right-movings (indistinguishability of left and right).

Besides the fuzziness of left- and right-movings another related feature that noncommutative chiral bosons possess is that $x^{+}$ and $x^{-}$ take values only on a series of discrete hyperbolas defined by the quantum condition (26), while for commutative chiral bosons they take continuous values on the whole $(x^{+},x^{-})$ plane. In fact, this is a kind of quantum phenomenon that is consistent with the noncommutative algebra (9). We may draw a figure of a series of discrete hyperbolas for a specific choice of $\theta$, for instance, ${\theta}=\frac{1}{2\pi}$. See Figure 1. In addition, we point out that ${D_{-}}^{\ast}{\phi}=0$ gives the same solution as $D_{-}{\phi}=0$. It is evident that the two equations of motion are equivalent to each other because of the reality of $\phi$. Let us, on the other hand, discuss from the point of view of their solutions. Although the general solution of the former is $f(x^{+}){\rm exp}(+i\frac{1}{2\theta}x^{+}x^{-})$ which has a positive phase factor instead of the minus in eq.(25), it reduces exactly to eq.(27) after the requirement of reality is considered and the quantum condition (26) is obtained. Thanks to the requirement of reality,
the solution of noncommutative chiral bosons simultaneously satisfies the equation of motion (15) satisfied by the noncommutative generalization of massless real scalars. In the commutative spacetime it is quite straightforward that the solution of ${\partial}_{-}{\phi}=0$ must be the solution of ${\partial}_{+}{\partial}_{-}{\phi}=0$. This result is a necessary condition for a consistent chiral boson theory {\em not only} in the commutative spacetime {\em but also} in the noncommutative one. By using this result eq.(24) reduces to ${D_{-}}^{\ast}{\lambda}_{+}=0$, which is the equation of motion satisfied by the auxiliary field ${\lambda}_{\mu}(x)$. We do not need to solve it because this auxiliary degree of freedom can be eliminated by constraints as will be demonstrated in detail in the next section.       
\begin{figure}
\begin{center}
\includegraphics[scale=0.4]{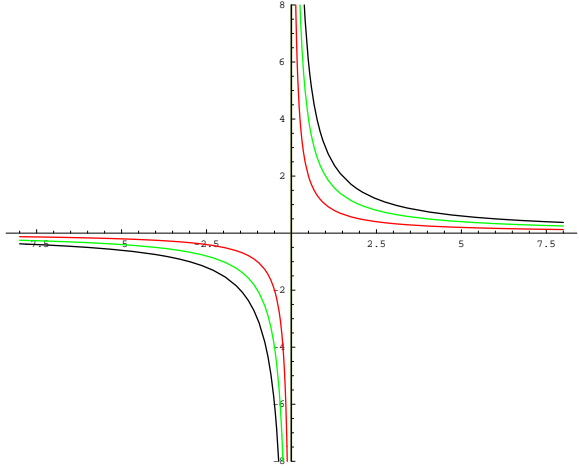}
\end{center}
\caption{Three pairs of discrete hyperbolas for a specific value of the noncommutative parameter, ${\theta}=\frac{1}{2\pi}$, are illustrated as examples, of which the red, green, and black curves correspond to the ``quantum number'' $n$ taking 1, 2, and 3, respectively. In general, $n$ may take any integer. For noncommutative chiral bosons described by the Lagrangian eq.(22), $x^{+}$ and $x^{-}$ take values only on the series of discrete hyperbolas governed by the ``quantum condition'' $x^{+}x^{-}=n$.}
\end{figure}

\section{Quantization of noncommutative chiral bosons}

Now we turn to quantize the constrained system described by eq.(22) and deal with a non-Hermitian Hamiltonian that occurs due to the specific Lagrangian of noncommutative chiral bosons. For the sake of convenience of making canonical analyses, we rewrite eq.(22) as 
\begin{eqnarray}
{\cal L}_{c}&=&\frac{1}{2}({\partial}_{0}{\phi})^{2}-\frac{1}{2}({\partial}_{1}{\phi})^{2}-\frac{1}{8{\theta}^2}[-(x^0)^2+(x^1)^2]{\phi}^2+i\frac{1}{2\theta}{\phi}(x^0{\partial}_{1}{\phi}+x^1{\partial}_{0}{\phi})\nonumber \\
& &+{\lambda}\left[i{\partial}_{0}{\phi}-i{\partial}_{1}{\phi}-\frac{1}{2\theta}(x^{0}+x^{1}){\phi}\right],
\end{eqnarray}
where ${\lambda} \equiv \frac{1}{\sqrt{2}}{\lambda}_{+}$. Note that imaginary terms appear in the Lagrangian and will also exist in the Hamiltonian even if we drop the fourth term which is a total derivative. This produces a seemingly thorny problem that the Hamiltonian is no longer Hermitian, which is a common problem in noncommutative field theories though the problem does not happen in the noncommutative real scalar field described by eq.(13) or eq.(17). In higher (than two) dimensional spacetime this problem may be circumvented by requiring a commutative time, but the requirement cannot be imposed on our case. However, the non-Hermiticity may not bring about a fatal difficulty. Based on the study of complex Hamiltonians~\cite{s20}, one may loosen the condition of Hermiticity and replace it by a weaker condition of spacetime reflection symmetry, {\em i.e.}, ${\cal PT}$ symmetry. In the following we will verify that the reduced Hamiltonian of the constrained system is ${\cal PT}$-symmetric and therefore the Lagrangian description (22) or (29) is physically acceptable though the non-Hermiticity is inevitable.

The momenta conjugate to ${\phi}$ and $\lambda$ are defined, respectively, by 
\begin{equation}
{\pi}_{\phi} \equiv {\partial}{\cal L}_{c}/{\partial}({\partial}_{0}\phi)={\partial}_{0}{\phi}+i\frac{1}{2\theta}x^1{\phi}+i{\lambda},
\end{equation}
and
\begin{equation}
{\pi}_{\lambda} \equiv {\partial}{\cal L}_{c}/{\partial}({\partial}_{0}\lambda)=0.
\end{equation}
Eq.(31) means a primary constraint which is expressed by
\begin{equation}
{\Omega}_{1}(x) \equiv {\pi}_{\lambda}(x) \approx 0,
\end{equation}
where $\approx$ stands for Dirac's weak equality. By means of the Legendre transformation we calculate the canonical Hamiltonian
\begin{eqnarray}
{\cal H}_{c}&=&{\pi}_{\phi}{\partial}_{0}{\phi}+{\pi}_{\lambda}{\partial}_{0}{\lambda}-{\cal L}_{c}\nonumber \\
&=&\frac{1}{2}\left({\pi}_{\phi}-i\frac{1}{2\theta}x^1{\phi}-i{\lambda}\right)^{2}+\frac{1}{2}({\partial}_{1}{\phi})^2+\frac{1}{8{\theta}^2}[-({x^0})^2+(x^1)^2]{\phi}^2-{i}\frac{1}{2\theta}x^0{\phi}{\partial}_{1}{\phi}\nonumber \\
& &+{\lambda}\left[i{\partial}_{1}{\phi}+\frac{1}{2\theta}(x^{0}+x^{1}){\phi}\right].   
\end{eqnarray}
As primary constraints should be preserved in time, which is treated as a basic consistency requirement in dynamics of constrained systems, we then have a secondary constraint
\begin{eqnarray}
{\Omega}_{2}(x) &=& \int dy^{1}\{\Omega_1(x), {\cal H}_c(y)\}_{PB}\nonumber \\
&=&i({\pi}_{\phi}(x)-{\partial}_{1}{\phi}(x))-\frac{1}{2\theta}x^{0}{\phi}(x)+{\lambda}(x) \approx 0, 
\end{eqnarray}
where $\{,\}_{PB}$ stands for an equal-time Poisson bracket.
No further constraints exist and the two constitute a second-class set. By using Dirac's method~\cite{s19}, we derive the following non-vanishing equal-time Dirac brackets:
\begin{eqnarray}
{\{{\phi}(x),{\pi}_{\phi}(y)\}}_{DB}&=&{\delta}(x^1-y^1),\nonumber \\
{\{{\phi}(x),{\lambda}(y)\}}_{DB}&=&-i{\delta}(x^1-y^1),\nonumber \\
{\{{\pi}_{\phi}(x),{\lambda}(y)\}}_{DB}&=&i{\partial}_{1}{\delta}(x^1-y^1)-\frac{1}{2\theta}x^{0}{\delta}(x^1-y^1),\nonumber \\
{\{{\lambda}(x),{\lambda}(y)\}}_{DB}&=&2{\partial}_{1}{\delta}(x^1-y^1).
\end{eqnarray}
In the sense of Dirac brackets weak constraints become strong conditions. As a consequence, we obtain the reduced Hamiltonian expressed in terms of independent degrees of freedom in phase space by solving $\lambda$ from eq.(34) and substituting into eq.(33)
\begin{equation}
{\cal H}_{r}={\pi}_{\phi}\left[{\partial}_{1}{\phi}+i\frac{1}{2\theta}(x_{0}-x_{1}){\phi}\right]. 
\end{equation}
As mentioned in section 3, the Lagrange multiplier $\lambda$ has been eliminated by constraints. Note that when the noncommutative algebra (9) tends to its full Heisenberg commutation relation, {\em i.e.}, ${\theta} \longrightarrow \infty$, the reduced Hamiltonian reduces exactly to the formulation of commutative chiral bosons as shown in Ref.~\cite{s14}. This guarantees the consistency of our generalization in one aspect from the point of view of quantization. The other aspect is shown by the fact that the Hamiltonian equation of motion for $\phi$ coincides with the self-duality condition, that is, 
\begin{equation}
{\partial}_{0}{\phi}(x)=\int dy^{1}{\{{\phi}(x),{\cal H}_{r}(y)\}}_{DB}={\partial}_{1}{\phi}(x)+i\frac{1}{2\theta}(x_{0}-x_{1}){\phi},
\end{equation}
is the same as $D_{-}{\phi}=0$. Moreover, we may compute the Hamiltonian equation of motion for ${\pi}_{\phi}$,
\begin{equation}
{\partial}_{0}{\pi}_{\phi}(x)=\int dy^{1}{\{{\pi}_{\phi}(x),{\cal H}_{r}(y)\}}_{DB}={\partial}_{1}{\pi}_{\phi}(x)-i\frac{1}{2\theta}(x_{0}-x_{1}){\pi}_{\phi},
\end{equation}
which may be rewritten in short by ${D_{-}}^{\ast}{\pi}_{\phi}=0$.

If we define ${\Pi}^{\mu} \equiv {\partial}{\cal L}_{c}/{\partial}({\partial}_{\mu}\phi)$, that is,
\begin{eqnarray}
{\Pi}^{0} &\equiv& {\partial}{\cal L}_{c}/{\partial}({\partial}_{0}\phi)={\partial}_{0}{\phi}+i\frac{1}{2\theta}x^1{\phi}+i{\lambda},\nonumber \\
{\Pi}^{1} &\equiv& {\partial}{\cal L}_{c}/{\partial}({\partial}_{1}\phi)=-{\partial}_{1}{\phi}+i\frac{1}{2\theta}x^0{\phi}-i{\lambda},
\end{eqnarray}
where ${\Pi}^{0}$ coincides with ${\pi}_{\phi}$ defined in eq.(30), we may calculate their light-cone components
\begin{eqnarray}
{\Pi}_{+} &\equiv& \frac{1}{\sqrt{2}}({\Pi}_{0}+{\Pi}_{1})=-iD_{+}{\phi}-i{\lambda}_{+},\nonumber \\
{\Pi}_{-} &\equiv& \frac{1}{\sqrt{2}}({\Pi}_{0}-{\Pi}_{1})=-iD_{-}{\phi}.  
\end{eqnarray}
The purpose of doing this is to re-formulate the Lagrangian eq.(22) and the equation of motion eq.(23) for noncommutative chiral bosons as follows:
\begin{equation}
{\cal L}_{c}={\Pi}_{+}{\Pi}_{-},  
\end{equation}
and
\begin{equation}
{\Pi}_{-}=0. 
\end{equation}
Alternatively, in a manifestly Lorentz covariant way the Lagrangian takes the form
\begin{equation}
{\cal L}_{c}=\frac{1}{2}{\Pi}_{\mu}({\eta}^{{\mu}{\nu}}+{\epsilon}^{{\mu}{\nu}})D_{\nu}\phi,
\end{equation}
where the ``covariant'' derivative is defined by
\begin{equation}
D_{\mu} \equiv -i{\partial}_{\mu}-\frac{1}{2\theta}{\epsilon}_{{\mu}{\nu}}x^{\nu}.
\end{equation}
$D_{\pm}$ and $D_{\mu}$ satisfy the usual relation between light-cone components defined by eq.(14) and Cartesian ones by eq.(44)
\begin{equation}
D_{+}=\frac{1}{\sqrt{2}}(D_{0}+D_{1}), \hspace{10mm}
D_{-}=\frac{1}{\sqrt{2}}(D_{0}-D_{1}).
\end{equation}
The formulation in this section is a natural generalization of that of chiral bosons in the ordinary spacetime given in Ref.~\cite{s14}.

After replacing ${\{,\}}_{DB}$ by $-i{[,]}_{ET}$ and $\phi$ and ${\pi}_{\phi}$ by their corresponding operators and making the product of operators symmetric in ${\cal H}_{r}$ in order to avoid operator ordering ambiguity, we therefore arrive at the quantized theory of noncommutative chiral bosons. The equal-time commutation relation of the operators of the independent degrees of freedom in phase space (${\phi},{\pi}_{\phi}$) takes the same form as eq.(20). We thus do not repeat it but just note that the symbols there should be understood as operators of the chiral boson field $\phi$ and its conjugate momemtum ${\pi}_{\phi}$ in this section. At present we may conclude that the Dirac method can be consistently applied to our noncommutative chiral boson model whose Lagrangian contains explicitly space and time.  

We now achieve the remaining task that is crucial to a physical model, that is, to verify that ${\cal H}_r$ possesses ${\cal PT}$ symmetry. In Ref.~\cite{s20} only systems of quantum mechanics in ordinary spacetime were discussed. Therefore, we have to extend the treatment there to field theory in noncommutative spacetime in our case. We begin with the ${\cal P}$ and ${\cal T}$ transformations for a quantum mechanics system in commutative spacetime
\begin{eqnarray}
{\cal P}: & & x^{0}\longrightarrow x^{0},\hspace{6mm}  x^{1}\longrightarrow -x^{1}; \hspace{6mm} p_{0}\longrightarrow p_{0},\hspace{6mm} p_{1}\longrightarrow -p_{1}, \nonumber \\
{\cal T}: & & x^{0}\longrightarrow -x^{0},\hspace{6mm}  x^{1}\longrightarrow x^{1};\hspace{6mm}  p_{0}\longrightarrow p_{0},\hspace{6mm}  p_{1}\longrightarrow -p_{1};\hspace{6mm}i{\longrightarrow}-i,  
\end{eqnarray}
each of which keeps the standard Heisenberg commutation relation (7) unchanged. In noncommutative spacetime we add such a transformation of the noncommutative parameter\footnote{A more general form of this transformation, ${\theta}_{0i}\longrightarrow -{\theta}_{0i}$ under parity was used~\cite{s21} to investigate ${\cal C}$, ${\cal P}$, and ${\cal T}$ invariance of noncommutative gauge theories based on the full Heisenberg commutation relation.}
under ${\cal P}$ and ${\cal T}$
\begin{equation}
{\cal P}: {\theta}\longrightarrow -{\theta}, \hspace{10mm}
{\cal T}: {\theta}\longrightarrow {\theta}, 
\end{equation}
that the resulting transformations of ${\hat x}^{\mu}$ and ${\hat p}_{\mu}$ maintain the noncommutative algebra (9) unchanged. 
Substituting eqs.(46) and (47) into the realization of the noncommutative algebra, eq.(8), we find that the transformations of ${\hat x}^{\mu}$ and ${\hat p}_{\mu}$ take the same form as $x^{\mu}$ and $p_{\mu}$, respectively, which, combined by eq.(47), can be written in a complete form
\begin{eqnarray}
{\cal P}: & & {\hat x}^{0}\longrightarrow {\hat x}^{0},\hspace{6mm}  {\hat x}^{1}\longrightarrow -{\hat x}^{1}; \hspace{6mm} {\hat p}_{0}\longrightarrow {\hat p}_{0},\hspace{6mm} {\hat p}_{1}\longrightarrow -{\hat p}_{1};\hspace{6mm} {\theta}\longrightarrow -{\theta}, \nonumber \\
{\cal T}: & & {\hat x}^{0}\longrightarrow -{\hat x}^{0},\hspace{6mm}  {\hat x}^{1}\longrightarrow {\hat x}^{1};\hspace{6mm}  {\hat p}_{0}\longrightarrow {\hat p}_{0},\hspace{6mm}  {\hat p}_{1}\longrightarrow -{\hat p}_{1}; \hspace{6mm}i{\longrightarrow}-i.  
\end{eqnarray}
It is easy to prove that the noncommutative algebra (9) is indeed invariant under each of the above transformations. This is a consistency condition for the spacetime reflection.
In addition, for field theory the ${\cal P}$ and ${\cal T}$ should include the transformations of the real scalar field $\phi$ and its conjugate momentum ${\pi}_{\phi}$ as follows:
\begin{eqnarray}
{\cal P}: & & {\phi}\longrightarrow {\phi}, \hspace{10mm}{\pi}_{\phi}\longrightarrow {\pi}_{\phi},\nonumber \\
{\cal T}: & & {\phi}\longrightarrow {\phi}, \hspace{10mm}{\pi}_{\phi}\longrightarrow -{\pi}_{\phi},  
\end{eqnarray}
which obviously keep the commutation relation (20) unchanged. By using eqs.(46), (47), and (49), we therefore verify that the reduced Hamiltonian (36) is ${\cal PT}$-symmetric, that is,
\begin{equation}
{\cal PT}{\cal H}_{r}={\cal H}_{r}.
\end{equation}
Note that ${\cal H}_r$ is not invariant under either ${\cal P}$ or ${\cal T}$, but invariant under the combination of them. As a consequence, the noncommutative linear constraint model of chiral bosons proposed in section 3 is physically acceptable even though its Hamiltonian is non-Hermitian.

\section{Non-self-duality of the noncommutative chiral boson action}

Recently the authors of Ref.~\cite{s22} obtained several new bosonic {\em p}-brane actions (including strings) with and without the Weyl-invariane by means of the systematic parent action approach and established duality symmetries in the set of known actions and of new ones as well. It may be interesting to investigate the duality symmetry for the newly proposed noncommutative chiral boson action in this paper. To this end, we write the action of eq.(22) in a manifestly Lorentz covariant form
\begin{equation}
S_{c}=\int d^{2}x\left[\frac{1}{2}D^{\mu}{\phi}D_{\mu}{\phi}+\frac{1}{2}{\lambda}^{\mu}\left(D_{\mu}{\phi}+{\epsilon}_{{\mu}{\nu}}D^{\nu}{\phi}\right)\right],  
\end{equation}
where $D_{\mu}$ is defined by eq.(44). In terms of the variation principle, we can calculate the classical equations of motion
\begin{equation}
({\eta}^{{\mu}{\nu}}+{\epsilon}^{{\mu}{\nu}})D_{\nu}{\phi}=0,
\end{equation}
and
\begin{equation}
{D^{\mu}}^{\ast}D_{\mu}{\phi}+\frac{1}{2}({\eta}^{{\mu}{\nu}}+{\epsilon}^{{\mu}{\nu}}){D_{\nu}}^{\ast}{\lambda}_{\mu}=0,
\end{equation}
which are equivalent to eq.(23) and eq.(24), respectively.

Introducing two auxiliary vector fields, $F_{\mu}$ and $G^{\mu}$, we construct the parent action that corresponds to the original action (51)
\begin{equation}
S_{\rm parent}=\int d^{2}x\left[\frac{1}{2}F^{\mu}F_{\mu}+\frac{1}{2}{\lambda}^{\mu}\left(F_{\mu}+{\epsilon}_{{\mu}{\nu}}F^{\nu}\right)+G^{\mu}\left(F_{\mu}-D_{\mu}{\phi}\right)\right].  
\end{equation}
Varying eq.(54) with respect to $G^{\mu}$ leads to 
\begin{equation}
F_{\mu}=D_{\mu}{\phi},
\end{equation}
together with which eq.(54) reduces to eq.(51). This shows the classical equivalence between the parent and original actions. On the other hand, varying eq.(54) with respect to $F_{\mu}$ gives the expression of $F^{\mu}$ in terms of $G^{\mu}$ and ${\lambda}_{\mu}$
\begin{equation}
F^{\mu}=-G^{\mu}-\frac{1}{2}\left({\lambda}^{\mu}-{\epsilon}^{{\mu}{\nu}}{\lambda}_{\nu}\right). 
\end{equation}
If we define ${\cal F}_{\mu} \equiv F_{\mu}+{\epsilon}_{{\mu}{\nu}}F^{\nu}$, we can write the self-duality condition in a simpler form: ${\cal F}_{\mu}=0$. If we apply a similar definition to $G^{\mu}$, such as ${\cal G}_{\mu} \equiv G_{\mu}+{\epsilon}_{{\mu}{\nu}}G^{\nu}$, by using eq.(56) we find the relation 
\begin{equation}
{\cal F}_{\mu}=-{\cal G}_{\mu}. 
\end{equation}
This means that the self-duality takes the same formula in terms of both $F_{\mu}$ and $G_{\mu}$ which are related with a generalized anti-dualization for commutative chiral bosons~\cite{s16} but without such an equation for noncommutative chiral bosons if the noncommutative parameter is finite. Substituting eq.(56) into eq.(54), we obtain the dual version of the original action
\begin{equation}
S_{\rm dual}=\int d^{2}x\left[-\frac{1}{2}G^{\mu}G_{\mu}-\frac{1}{2}{\lambda}^{\mu}\left(G_{\mu}+{\epsilon}_{{\mu}{\nu}}G^{\nu}\right)-{\phi}{D_{\mu}}^{\ast}G^{\mu}\right].  
\end{equation}
The variation of eq.(58) with respect to $\phi$ treated at present as a Lagrange multiplier leads to the equation, 
\begin{equation}
{D_{\mu}}^{\ast}G^{\mu}=0. 
\end{equation}
Till now every step of the parent action approach is exactly the same as that for ordinary chiral bosons. The difference appears when we try to solve the equation. If ${\theta}\longrightarrow \infty$, the solution of eq.(59) is proportional to ${\epsilon}^{{\mu}{\nu}}{\partial}_{\nu}{\varphi}$, where $\varphi$ is an arbitrary real function, and $S_{\rm dual}$ reduces to the original action in terms of ${\varphi}$ which relates to ${\phi}$ with a generalized anti-dualization~\cite{s16}. However, for a finite $\theta$ we find that the solution of eq.(59) must {\em not} be proportional to ${\epsilon}^{{\mu}{\nu}}{D}_{\nu}{\varphi}$ because ${D_{\mu}}^{\ast}D_{\nu}$ or $D_{\mu}{D_{\nu}}^{\ast}$ is not symmetric under the permutation of lower indices $\mu$ and $\nu$. On the other hand, this kind of solution is indispensable in order for the dual action to take the same form as the original action, in other words, for the noncommutative chiral boson action to possess self-duality. As a consequence, we conclude that the self-duality does not exist in our noncommutative chiral boson action but does exist in the commutative case.

\section{Conclusion}

What we have done in the above sections can be summarized as follows. As a basis we choose a noncommutative algebra with noncommutativity of both coordinates and momenta. Simulating the procedure of establishing field theory from mechanics, we then obtain three different kinds of noncommutative generalizations of massless real scalar fields after finding a suitable realization to the noncommutative algebra. One feature of the realization is the mixture between the noncommutativity of coordinates and that of momenta. Namely, with more noncommutativity of 
coordinates there is less of momenta
and {\em vice versa}.
In ordinary spacetime the Lagrangian of massless real scalar fields can be written as the product of left- and right-handed chiralities. In noncommutative spacetime, the result is needed but should be extended to a more general form. We therefore select only one from the three different formulations mentioned above that can, as we desire, be expressed as the product of generalized left- and right-handed chiralities. 
After introducing a Lagrange multiplier we obtain the Lagrangian for noncommutative chiral bosons by linearly imposing the generalized left- or right-handed chirality on the Lagrangian of massless real scalar fields in a manifestly Lorentz covariant way. An interesting ambiguity of left- and right-moving chiral bosons can thus be deduced from the solution of the equation of motion of noncommutative chiral bosons. This fuzzy phenomenon, we may call it with such a name, originates from no other source but the noncommutativity of coordinates and momenta for a classical theory. We note that the requirement of reality of chiral boson fields plays a crucial role {\em not only} in exposing the so-called fuzziness {\em but also} in keeping the theory consistent. This was followed by the Dirac quantization of the constrained system we proposed. We find that Dirac's method is smoothly extended to our system that contains explicitly space and time. In particular, the Hamiltonian of the system is not Hermitian but, fortunately, is ${\cal PT}$-symmetric that is a less restrictive condition. Finally, no self-duality exists in the noncommutative chiral boson action because of the appearance of the noncommutative parameter.    

For further development one aspect is to consider the noncommutative generalization for chiral {\em p}-forms ($p \geq 1$) as mentioned in footnote 4. Some interesting noncommutative phenomena, similar to the fuzzy left- and right-movings for chiral bosons, may occur. It may be an appealing topic in connecting string theory to noncommutativity from the point of view that is different from the one already proposed in Ref.~\cite{s9} because chiral {\em p}-forms have a close relationship with string theory. Another aspect of our considerations now is still to construct noncommutative models for the other chiral boson theories with or without the manifest Lorentz covariance, which is a straightforward extension of the present work. For instance, one may suggest a noncommutative formulation of the Siegel model~\cite{s13} with such an   
equation of motion for chiral boson fields, $(D_{-}{\phi})^2=0$. The classical solution is the same as that of eq.(23). That is, the same fuzzy phenomenon as mentioned above may exist. However, at the quantum level the squared form is quite different from the linear one, the former is a first class constraint while the latter is second class. This brings about a completely different constraint structure. Possible new results related to the difference are now being studied. Moreover, in the commutative spacetime, as we know, the Siegel model changes to Floreanini-Jackiw's~\cite{s12} which is not manifestly Lorentz covariant if an additional constraint is imposed~\cite{s23}. Similar treatment may also be applied in the noncommutative spacetime and one possible noncommutative generalization of the Floreanini-Jackiw model may be obtained through imposing a suitable constraint upon the noncommutative generalization of Siegel's model. Details of analyses will be reported separately later.

\vspace{8mm}
\noindent
{\em Note added}: Following the noncommutative generalization of
chiral bosons proposed in this paper, we
deduce several technical results that do not appear inconsistent at
the classical level such as, for instance, the discovery of
a ``fuzzy'' spatial dimension and the non-existence of self-duality. Moreover,
we also perform formally the canonical
quantization of noncommutative chiral bosons. However, from the point of
view of some requirements that a physically
well-defined quantum theory should satisfy, our
noncommutative model is non-positive definite and
involves a non-Hermitian Hamiltonian. Because the linear constraint model
of chiral bosons in ordinary
spacetime~\cite{s14} is intrinsically non-positive definite, the
non-positive definition here should not be induced
by our noncommutative generalization but originates from the corresponding
commutative formulation. As supposed in
the last section, nonperturbative instabilities caused by this non-positive
definition may be circumvented by
constructing a possible noncommutative generalization of the
Floreanini-Jackiw model whose positive definition is obvious in ordinary spacetime. As to the non-Hermitian
Hamiltonian that includes a noncommutative parameter, it appears indeed due
to our generalization to noncommutative
spacetime. In general, an additional phase factor that is closely related to
noncommutative parameters usually
occurs and probably gives rise to a non-Hermitian Hamiltonian in 
noncommutative field theories with space-time
noncommutativity. A non-Hermitian Hamiltonian leads to non-unitary
evolution. For the sake of consistency, extra
fields should appropriately be introduced in the traditional treatment,
which may change the Hamiltonian to be
positive definite and/or Hermitian. Alternatively, one may try to
construct~\cite{s20} a unitary operator relating a
non-Hermitian Hamiltonian with exact ${\cal PT}$ symmetry to a Hermitian
Hamiltonian and may establish certain
relations\footnote{It is claimed in the third citation of Ref.~\cite{s20} that an
equivalence between the exact ${\cal PT}$ symmetry and Hermiticity can be
established. However, this equivalence
is realized only in the two dimensional
Hilbert space and may probably be extended to a higher (than two) but finite dimensional
Hilbert space.} between the two types
of Hamiltonians. This unitary operator might be available since the ${\cal PT}$ symmetry exists in our case.
However, the price to be paid for replacing the Hermiticity condition by the exact
${\cal PT}$ symmetry in an infinite dimensional Hilbert space probably gives
rise to non-locality of field theories. It might be interesting to compare
the non-locality caused by this
replacement with that by the generalization of commutative to noncommutative
spacetime in noncommutative field
theories. We discuss in this paper a model of field theories with space-time
noncommutativity, which perhaps goes
beyond the results of Ref.~\cite{s24}, that is, there does not appear to
exist a decoupling limit within string
theory that isolates purely field theoretic degrees of freedom with
space-time noncommutativity (in Lorentzian
signature). Something interesting might be to study the extension of our
discussions to Euclidean signature (possibly thinking of holomorphicity as
the analog of chirality), where they
might be interpreted in terms of a
statistical mechanical system and the conceptual problems of time are
absent. We thank the referee for helpful comments.

\newpage
\noindent
{\bf Acknowledgments}
\vspace{5mm}

\noindent
Y.-G. M. would like to thank Nobuyoshi Ohta for helpful discussions and comments, and to acknowledge the support through him from the Grant-in-Aid for Scientific Research. He also thanks Tetsuo Shindou for the help of drawing a figure. This work was supported in part by 
the National Natural Science Foundation of China under grant No.10275052.
D.K.P. acknowledges support by the Korea Research Foundation under
Grant (KRF-2002-015-CP0063).

\end{document}